\documentclass[showpacs,prl,onecolumn,aps,superscriptaddress,preprintnumbers,nofootinbib]{revtex4}
\usepackage{amsmath,amssymb}
\usepackage{epsfig}
\usepackage{graphicx}
\usepackage{amsmath}
\usepackage{amsfonts}
\usepackage{epstopdf}
\def\slashchar#1{\setbox0=\hbox{$#1$}     		% set a box for #1
   \dimen0=\wd0                                 	% and get its size
   \setbox1=\hbox{/} \dimen1=\wd1               	% get size of /
   \ifdim\dimen0>\dimen1                        	% #1 is bigger
      \rlap{\hbox to \dimen0{\hfil/\hfil}}      	% so center / in box
      #1                                        	% and print #1
   \else                                        	% / is bigger
      \rlap{\hbox to \dimen1{\hfil$#1$\hfil}}   	% so center #1
      /                                         	% and print /
   \fi}

\renewcommand{\vec}{\boldsymbol}
\newcommand{\beq}{\begin{equation}}
\newcommand{\eeq}{\end{equation}}
\newcommand{\bea}{\begin{eqnarray}}
\newcommand{\eea}{\end{eqnarray}}
\newcommand{\ba}{\begin{array}}
\newcommand{\ea}{\end{array}}

\def\eq#1{{Eq.~(\ref{#1})}}
\def\fig#1{{Fig.~\ref{#1}}}
\newcommand{\bas}{\bar{\alpha}_S}
\newcommand{\as}{\alpha_S}
\newcommand{\nn}{\nonumber}

\newcommand{\h}{\frac{1}{2}}

\newcommand{\ha}{{\cal H}}

\newcommand{\Lb}{\left(}
\newcommand{\Rb}{\right)}

\begin{document}

\title{ A new parton model for the soft interactions at high energies.	}
\author{E. ~Gotsman}
\email{gotsman@post.tau.ac.il}
\affiliation{Department of Particle Physics, School of Physics and Astronomy,
Raymond and Beverly Sackler
 Faculty of Exact Science, Tel Aviv University, Tel Aviv, 69978, Israel}
 \author{ E.~ Levin}
\email{leving@tauex.tau.ac.il, eugeny.levin@usm.cl}
\affiliation{Department of Particle Physics, School of Physics and Astronomy,
Raymond and Beverly Sackler
 Faculty of Exact Science, Tel Aviv University, Tel Aviv, 69978, Israel}
 \affiliation{Departemento de F\'isica, Universidad T\'ecnica Federico
 Santa Mar\'ia, and Centro Cient\'ifico-\\
Tecnol\'ogico de Valpara\'iso, Avda. Espana 1680, Casilla 110-V,
 Valpara\'iso, Chile} 
 \author{  I.~ Potashnikova}
\email{irina.potashnikova@usm.cl}
\affiliation{Departemento de F\'isica, Universidad T\'ecnica Federico
 Santa Mar\'ia, and Centro Cient\'ifico-\\
Tecnol\'ogico de Valpara\'iso, Avda. Espana 1680, Casilla 110-V,
 Valpara\'iso, Chile}
\date{\today}


\begin{abstract}

 We  propose a new parton model and demonstrate that 
the model describes the relevant
 experimental data at high energies. The 
model
 is based on  Pomeron calculus in 1+1 space-time dimensions, as suggested 
in
 Ref. \cite{KLL}, and on  simple assumptions regarding the hadron
 structure, related to the impact parameter dependence of the
 scattering amplitude. This parton model  evolves from QCD, assuming
 that the unknown non-perturbative corrections lead to fixing  the
 size of the interacting dipoles. The advantage of this approach is
 that it satisfies both t-channel and s-channel unitarity, and
 can be used for summing all diagrams of  Pomeron interactions,
 including  Pomeron loops.  We can use this
 approach for all  reactions: dilute-dilute (hadron-hadron),
 dilute- dense (hadron - nucleus) and dense-dense (nucleus-nucleus)
for the scattering of
 parton systems.  Unfortunately, we are still far  from
 being  able to tackle  this problem in the  effective QCD  
theory   at high energy  (i.e.  in the CGC /saturation approach).
 \end{abstract}
 
 \preprint{TAUP-}

\maketitle

\tableofcontents

\flushbottom

\section{Introduction}
 In our previous papers \cite{GLP1,GLP2,GLMNI,GLM2CH,GLMINCL,GLMCOR,GLMSP,
GLMACOR} we demonstrated that it is possible to build a model, based on the
 effective QCD theory at high energies: i.e. Colour Glass Condensate
(CGC) approach (see Ref.\cite{KOLEB} for the review).  The success of
 the model  emanates from two principle ideas, which  are  
supported  
 by experimental data: (i)  typical distances in the soft processes
 at high energies, turn out to be rather short; and (ii) the CGC approach
 can be re-written in an equivalent form, as the interaction of  BFKL
 Pomerons\cite{AKLL} in a limited range of rapidities
 ( $Y \leq Y_{\rm max}$):

 \beq \label{RAPRA}
Y \,\leq\,\frac{2}{\Delta_{\mbox{\tiny BFKL}}}\,\ln\Lb
 \frac{1}{\Delta^2_{\mbox
{\tiny BFKL}}}\Rb
\eeq 
 where $\Delta_{\mbox{\tiny BFKL}}$ denotes the intercept of the BFKL 
 Pomeron\cite{BFKL}. In our model $ \Delta_{\mbox{\tiny BFKL}}\,
\approx\,0.2 - 0.25$    leading to $Y_{max} = 20 - 30$, which covers
 all collider energies.     

 The equivalence between the CGC approach and the BFKL Pomeron calculus
 is very important, since (i) it  shows that the CGC approach satisfies
  t-channel unitarity, which is  a prerequisite for any effective 
theory at
 high energies\footnote{We wish to remind  the reader that  
t-channel unitarity
 follows from two sources: the t-channel unitarity of the BFKL Pomeron,
 proven in Ref.\cite{BFKL}, and from the Gribov Reggeon Diagram
 Technique\cite{GRIBRC}.} and (ii) it allows us to consider 
 elastic and diffraction processes, and  processes of multiparticle
 generation, on the same footing.  The last statement follows
 from the AGK cutting rules\cite{AGK}, which has been proven for 
  inclusive  production\cite{KOTU},  but not for 
 correlations  \cite{KOJA}.

 In Ref.\cite{AKLL} it was shown than in the rapidity range of 
\eq{RAPRA} we
 can use the MPSI approximation\cite{MPSI}, which sums the large Pomeron 
loops
 for the  following Hamiltonian
\begin{eqnarray}\label{HB}
\ha&=&\frac{N_c^2}{2\pi\bar\alpha_s}\int\bar P(x,y)\nabla^2_x\nabla^2_y\left[K(x,y|z)P(x,z)+P(z,y)- P(x,y)- P(x,z)P(z,y)\right]\nn\\
&  & - P(x,y)\nabla^2_x\nabla^2_y\left[
K(x,y|z)\bar P(x,z)\bar P(z,y)\right]
\end{eqnarray}
where $$K(x,y|z)=\frac{(\vec{x} -\vec{y})^2}{(\vec{x} - \vec{z})^2(\vec{y} - \vec{z})^2},$$  $P$ and $\bar P$ 
denote
 the BFKL Pomeron fields.

The commutation  relations are of the form
\beq \label{CR}
[\bar P(x,y), P(u,v)]\,=\,\gamma(x,y; u,v)
\eeq 
 where $\gamma(x,y; u,v)$ is the scattering amplitude of a dipole $(x,y)$ on
 the dipole $(u,v)$
\beq\label{DDAM}
 \gamma(x,y; u,v)=\frac{\alpha_s^2}{32\pi^2}\ln^2\frac{(\vec{x} -  \vec{u})^2(\vec{y} -\vec{v})^2}{(\vec{x} - \vec{v})^2(\vec{y} - \vec{u})^2}
\eeq 
 
 The above form of the Hamiltonian was suggested  in Ref.\cite{BRAUN} as 
the
 natural way for  summing all  BFKL Pomeron diagrams with only  triple
 Pomeron interactions.  Below, we will denote this Hamiltonian by 
$\ha_{\rm B}$.  
 
In Ref.\cite{KLL} it is shown that the Hamiltonian of \eq{HB} cannot be
  the correct one, since it violates  s-channel unitarity.
 In other words, we can use $\ha_{\rm B}$ only for summing the large
 Pomeron loops in the MPSI approximation,  but cannot  use it  for a 
general
 description of the high energy interaction in QCD,   nor even for 
the
 description of the DIS, which is given by Balitsky - Kovchegov
 equation\cite{BK},  and  was considered  to be  the  most 
reliable
 equation in the framework of the CGC approach. It should be stressed, 
that in
 the CGC approach, we do not have a general Hamiltonian that describes the
 interaction of two dense, or of two dilute systems of partons,  although 
some
 work in this direction has been done\cite{FOAM}. Unfortunately, 
 no progress in formulating the BFKL Pomeron calculus for these systems  has
 been achieved in Ref.\cite{KLL}.
 
 However, in Ref.\cite{KLL} such a Hamiltonian is constructed for the
 1+1 Reggeon Field Theory, which corresponds to  QCD in which the
 size of the interacting dipoles is fixed\cite{MUDI,LELU}. 
 In this model the BFKL equation for the dipole scattering cross section
 $\sigma$ at a rapidity $Y$ is reduced to
\beq \label{TM1}
\frac{d \sigma\Lb Y \Rb}{d Y} \,\,=\,\,\Delta  \, \sigma\Lb Y \Rb ,
\eeq
where $\Delta$ denotes the BFKL intercept. The 
\eq{TM1} reproduces the power-like increase of the cross section with
 energy, $\exp(\Delta Y) = (1/x)^\Delta$.  
For the scattering amplitude $N$ we obtain the non-linear equation:
\beq \label{N}
\frac{d N\Lb Y \Rb}{d Y} \,\,=\,\,\Delta  \,\Bigg( N\Lb Y \Rb\,\,-\,\,N^2\Lb Y\Rb\Bigg)
\eeq
 this   form is  similar to that of the Balitsky-Kovchegov non-linear
 equation\cite{BK}.
Note that  \eq{N} does not depend on the impact parameter . This
 fact is also in  agreement with the BFKL approach, in which the
 average momentum of the produced colourless dipoles increases with
 energy since $p_T \propto\,s^\lambda$, and  Gribov 
diffusion\cite{GRIBDIF}
 in impact parameter with $\Delta b   \propto \frac{1}{p_T} n\,=\,
 \frac{1}{p_T} \Delta Y$, leads to $\Delta b \to 0$ at high energies. 
  We can consider this simple model as the QCD approach in which
 the typical size of the colourless dipoles are fixed, and do not depend
 on energy. In other words,  we can view this simple approach as a new
 parton model for the high energy interaction\cite{PARTMOD}.
 
 In this model we  encountered the same problems with $s$-channel 
unitarity
 as in QCD at high energy, however we  have identified  the Pomeron 
Hamiltonian which
 cures all these problems\cite{KLL}. In the next section we will briefly 
review
  our finding. In section 3 we  propose a model based on this
 Hamiltonian, while in section 4  we compare the predictions of this 
model 
with the relevant
 experimental data.  We summarize our results in the Conclusions.

 %%%%%%%%%%%%%%%%%%%%%%%%%%%%%%%%%%%%%%%%%%%%%%%%%%%%%%
 \section{A new parton model}
 
 %%%%%%%%%%%%%%%%%%%%%%%%%%%%%%%%%%%%%%%%%%%%%%%%%%%%%%
 Below we give   a short review of  sections 4 and 5 of 
Ref.\cite{KLL}, where
  1+1 Reggeon Field Theory(RFT) is discussed. We emphasis  
 section 5, which contains the
 results that we are going to use in building our model.
 %%%%%%%%%%%%%%%%%%%%%%%%%%%%%%%%%%%%%%%%%%%%%%%%%%%%%%
 
 \subsection{Violation of the s-channel unitarity in 1 + 1 RFT}
 %%%%%%%%%%%%%%%%%%%%%%%%%%%%%%%%%%%%%%%%%%%%%%%%%%%%%%
 
 As a direct generalization of the  original QCD to 1+1 RFT,
the scattering matrix of the projectile consisting of $m$ dipoles on a
 target consisting of $\bar n$ dipoles, is given as by 
 \beq\label{AMP} 
\langle m|\bar n\rangle=\int d\bar P\delta(\bar P)(1-P)^m(1-\bar P)^{\bar n}
\eeq
The evolution in rapidity takes the form
\beq \label{EVRAPAMP}
\langle m|\bar n\rangle_Y=\int d\bar P\delta(\bar P)(1-P)^me^{HY}(1-\bar P)^{\bar n}
\eeq

The 1+1  analog of the BK evolution (see \eq{N})  is given by the Hamltonian
\beq\label{HBK}
\ha_{\rm BK}=-\frac{1}{\gamma}\Big(\bar PP-\bar PP^2\Big)
\eeq
As previously, we take $P$ and $\bar P$ to  satisfy the dilute limit 
algebra
 with the commutator \eq{CR}, such that
\beq \label{PBPDILI}
P\,=\,-\gamma\frac{d}{d\bar P}; \ \ \ \ \ \gamma \,\sim \,\bas^2 >0~~~~\leftarrow~~~\mbox{\eq{DDAM} in 1+1 RFT}
\eeq

To   comprehend  the origin of  $s$-channel unitarity,
 we  consider
 the infinitesimal evolution ($\delta Y$) of the projectile and target wave
 functions with the Hamiltonian $\ha_{\rm BK}$:
\begin{equation}\label{PROJEV}
\langle m|e^{\ha_{\rm BK}\delta Y}\,\,\propto\,\,  (1\,\,-\,\, m\,\delta Y )\langle m|\,\,+ \,\,m\,\delta Y\, \langle m+1| \end{equation}
 \begin{equation}\label{TAREV}
  e^{\ha_{\rm BK}\, \delta Y}|\bar n\rangle=(1\,+\, \bar n\,\delta Y)|\bar n\rangle \,\,-\,\,\bar n[1+\gamma(\bar n-1)]\,\delta Y|\bar n-1\rangle\,\,+\,\,\gamma\bar n(\bar n-1)\,\delta Y|\bar n-2\rangle
 \end{equation}
  Note the difference between these two equations. 
  The coefficient in front of an $n$-dipole state has the meaning of, the
 probability to find this number of dipoles in  the wave function. The
 projectile evolution, given by \eq{PROJEV}  is unitary: all the
 probabilities in the evolved state remain positive and smaller than
 unity, and the sum of the probabilities add up to unity.
 
  On the other hand, the target evolution is non-unitary.  Indeed,
 we face two difficulties with the unitarity: (i)
  the probability to find the initial state $|\bar n\rangle$ after
 a short interval of evolution exceeds unity; and (ii)  the probability
 to find a state $|\bar n  -  1\rangle$ is negative. The coefficients still
 sum to unity as for the projectile, but clearly  the target evolution
 violates unitarity.

 The Braun Hamiltonian of \eq{HB} takes the following form in 1+1 RFT:
 \begin{equation}\label{11HB}
\ha_{\rm B}\,\,=\,\,-\frac{1}{\gamma}\left[\bar PP-\bar PP^2-\bar P^2P\right]
 \end{equation}
 
  Exploring the same question about $s$-channel unitarity as in 
\eq{PROJEV} 
and \eq{TAREV} we obtain 
for the evolution of the target wave function:

\begin{equation}\label{TARHBEV}
e^{ \ha_{\rm B} \,\delta Y }|\bar n\rangle\,\,\approx\,\, (1+\ha_{\rm B}\delta Y )|\bar n\rangle\,\,=\,\,(1-  \bar n \delta Y)|\bar n\rangle +\bar n\,\delta Y|\bar n+1\rangle - \gamma\bar n(\bar n-1)\delta Y|\bar n-1\rangle + \gamma\bar n(\bar n-1)\delta Y|\bar n-2\rangle
\end{equation} 
One can see  that \eq{TARHBEV} violates unitarity,  but  this violation
  is smaller than in \eq{TAREV},  being $O(\gamma)$, and it is  small
 for small $\bar n$ . However, the coefficient of the term
 $|\bar n\rangle$ is still negative, and becomes  large 
 parametrically, long before the saturation limit is reached.
  Since \eq{11HB} is symmetric between the target and the projectile,
 the projectile evolution now is also non-unitary, and involves negative
 probabilities. The fact that the violation of the unitarity occurs at
 large $\bar n$  suggets that one reconsiders the commutation 
 relations of \eq{CR} and \eq{PBPDILI}, which we discuss in the
 next subsection.

 %%%%%%%%%%%%%%%%%%%%%%%%%%%%%%%%%%%%%%%%%%%%%%%%%%%%%%
 
 \subsection{Commutators}
 %%%%%%%%%%%%%%%%%%%%%%%%%%%%%%%%%%%%%%%%%%%%%%%%%%%%%%
 Using the commutators of \eq{CR} and \eq{PBPDILI}, the scattering matrix
 can be calculated explicitly
(for $m+1<\bar n$)
\beq \label{SMWR}
\langle m|\bar n\rangle\,=\,\sum_{l=0}^m\frac{m!\bar n!}{(m-l)!\,(\bar n-l)!\,l!}(-\gamma)^l
\eeq
In particular
\beq\label{}
\langle 1|\bar n\rangle=1-\bar n\gamma
\eeq

 From the above   one can see that the amplitudes become negative
 for $ \bar n \,>\,1/\gamma$,  and 
when a single dipole of the projectile scatters on several dipoles of the 
target,
our commutation relation does not allow us to account for multiple 
scattering corrections.

Therefore,  as well as  curing  the problems with unitarity,  we 
need to
 change \eq{CR}(\eq{PBPDILI}) in a such way that the scattering matrix will
 have the form:
\begin{equation} \label{CORAMP}
\langle 1|\bar n\rangle=\sum_{k=0}^{\bar n}\frac{\bar n!}{(\bar n-k)!k!}(-\gamma)^k
\end{equation}

 so as to correctly account for multiple rescatterings.

In Ref.\cite{KLL} the following   commutator  which reproduce
 \eq{CORAMP} is proposed.
\beq\label{CRCOR}
\Big(1\,\,-\,\,P\Big)\Big(1\,\,-\,\,\bar P \Big)\,\,=\,\,(1-\gamma)\Big(1\,\,-\,\,\bar P\Big) \Big(1\,\,-\,\,P\Big)
\eeq
\eq{CRCOR}
  gives the correct factor $(1-\gamma)^{\bar n}$ that includes
 all multiple scattering corrections, while all the dipoles remain intact,
 and can subsequently scatter on additional projectile or target dipoles.
For small $\gamma$,  and in the regime where  $P$ and $\bar P$ are 
also small, we obtain
\beq
[P,\bar P]=-\gamma +...
\eeq
consistent with our original expression. 

Note, that the algebra  of \eq{CRCOR} is equivalent to the following 
representation
\beq \label{defin} 1-\bar P=e^{ -\ln(1-\gamma)\frac {d}{d\Phi}}, ; \ \ \ \ \ \ 1-P=e^{-\Phi}
\eeq
where $\Phi $ and $\bar{\Phi}$ have the same meaning as in \eq{HB}.

In the calculation of an amplitude of the type of \eq{CORAMP}, once
 all the factors of $1-\bar P$ are commuted through to the left, 
then  in
 the remaining  matrix element $\bar P$ operates on the $\delta$-function 
and 
thus vanishes.
The remaining factors of $(1-P)$ also  become  unity, since a  factor
 of $\Phi$ is equivalent to a derivative acting on the $\delta$-function,
 and  vanishes when integrated over $\bar P$.

 With the new algebra we have
\begin{equation}\label{FINAMP}
\langle m|\bar n\rangle=(1-\gamma)^{m\bar n}
\end{equation}
which is a simple and intuitive result: the S-matrix of dipole-dipole
 scattering  raised to the power of the number of dipole pairs that 
scatter.

It should be stressed,  that the modification of  the Pomeron algebra
 is not a matter of choice, but is necessary to obtain the amplitude
  of \eq{CORAMP}, which is unitary for arbitrary numbers of colliding
 dipoles. However, the question of the unitarity of the evolution is a
 completely separate one.  We will examine the s-channel unitarity for
 the Braun Hamiltonian, since we plan to use this Hamiltonian for the
 description of hadron-hadron scattering at high energy.

We write the Braun Hamiltonian in a more convenient form:
\beq \label{HBB}
\ha_{\rm B}\,\,=\,\,-\frac{1}{\gamma}\Big[(1-\bar P)P-(1-\bar P)^2P+(1-\bar P)P^2-P^2\Big]
\eeq
The action on the projectile and the target is obviously symmetric, as
 the Hamiltonian is self dual under the transformation $P\rightarrow \bar P$.
\beq
e^{ \ha_{ \rm B}\,\delta Y}|\bar n\rangle=\left[1+\frac{\delta Y}{\gamma}\left[1-(1-\gamma)^{\bar n}\right]^2\right]|\bar n\rangle-\frac{\delta Y}{\gamma}\left[1-(1-\gamma)^{\bar n}\right]\left[2-(1-\gamma)^{\bar n}\right]|\bar n+1\rangle+\frac{\delta Y}{\gamma}\left[1-(1-\gamma)^{\bar n}\right]|\bar n+2\rangle
\eeq
\beq
\langle m|e^{ \ha_{ \rm B}\,\delta Y}=\left[1+\frac{\delta Y}{\gamma}\left[1-(1-\gamma)^{m}\right]^2\right]\langle m|-\frac{\delta Y}{\gamma}\left[1-(1-\gamma)^{m}\right]\left[2-(1-\gamma)^{m}\right]\langle m+1|+\frac{\delta Y}{\gamma}\left[1-(1-\gamma)^{m}\right]\langle m+2|
\eeq

This  result is a disaster:  the evolution of both, projectile and 
target are
 non-unitary. In fact the lack of unitarity  occurs  for an arbitrary
 number of dipoles $m$ and $\bar n$. It is shown in Ref.\cite{KLL}
 that the terms that include the four Pomeron interaction, do not
help.

%%%%%%%%%%%%%%%%%%%%%%%%%%%%%%%%%%%%%%%%%%%%%%%%%%%%%%
 
 \subsection{The Hamiltonian of the new parton model}
 %%%%%%%%%%%%%%%%%%%%%%%%%%%%%%%%%%%%%%%%%%%%%%%%%%%%%%
 In Ref.\cite{KLL} a new Hamiltonian is  proposed 
 
 \begin{equation}\label{HNPM}
\ha_{\rm NPM}=-\frac{1}{\gamma}\bar PP\eeq
where NPM stands for ``new parton model''.
The fact that it is self dual is  obvious.   In the
 limit of small $\bar P$ this Hamiltonian reproduces $\ha_{\rm BK}$
( see Ref.\cite{KLL} for details). This condition is  important
  for fixing the form of
$\ha_{\rm NPM}$.

To check  unitarity we consider:
\beq \label{UHNPM}e^{ \ha_{\rm NPM}\,\delta Y}|\bar n\rangle=\left[1-\frac{\delta Y}{\gamma}[1-(1-\gamma)^{\bar n}]\right]|\bar n\rangle+\frac{\delta Y}{\gamma}[1-(1-\gamma)^{\bar n}]|\bar n+1\rangle
\eeq
Performing the operation in \eq{UHNPM}, we 
first  express the $P^\dagger$ in terms of $\bar P$.
To do this, recall that $P^\dagger$ should annihilate a dipole when acting 
on
 the wave function. Using eq.(\ref{defin}) we can write
\beq \label{daggerbar}
P^\dagger= \frac{d}{d\Phi} e^{\Phi}=\frac{1}{\gamma}\ln(1-\bar P)\frac{1}{1-P}; \ \ \ \ \ \  \bar P^\dagger=-\frac{1}{\gamma} e^{-\gamma\frac{d}{d\Phi}}\Phi=\frac{1}{\gamma}\frac{1}{1-\bar P}\ln(1-P)\eeq
 For simplicity, in the above equations we have
 used $\ln(1-\gamma)\approx -\gamma$, since $\gamma\sim \alpha_s^2\ll 1$

To  check for unitarity, we see that
\beq \label{utmu}e^{ \ha_{\rm NPM}\,\,\delta Y}|\bar n\rangle=\left[1-\frac{\delta Y}{\gamma}[1-(1-\gamma)^{\bar n}]\right]|\bar n\rangle+\frac{\delta Y}{\gamma}[1-(1-\gamma)^{\bar n}]|\bar n+1\rangle
\eeq
showing that this evolution is clearly unitary. Due to self duality, it is 
clear
 that the evolution of the projectile wave function is unitary as well.  

It is interesting that \eq{utmu} displays
  saturation behavior very similar to that  expected from
  QCD evolution, namely at large $\bar n$, the change in the wave
 function is independent of the number of dipoles $\bar n$. In the BK
 approach in QCD, the wave function never saturates(see \eq{PROJEV}), and
  saturation of the scattering amplitudes  is due to  multiple
 scattering effects.
 
 %%%%%%%%%%%%%%%%%%%%%%%%%%%%%%%%%%%%%%%%%%%%%%%%%%%%
 \subsection{Equations of motion and the scattering amplitude }
%%%%%%%%%%%%%%%%%%%%%%%%%%%%%%%%%%%%%%%%%%%%%%%%%%%%%%
 The general form of the  equation of motion follows from
 \beq \label{EM1}
 \frac{d P}{d  \eta}\,=\,\Big[ \ha, P\Big]\,;~~~~{\rm and}~~~~ 
 \frac{d \bar P}{d  \eta}\,=\,\Big[\ha, \bar P\Big]
\eeq

For the Hamiltonian $\ha_{\rm NPM}$ we get 
\beq \label{H01}
\frac{ d P}{d  \eta}\,=\,\Lb 1 - \bar P\Rb \,\Lb 1 - P \Rb \,P;~~~~~~~\frac{ d \bar P}{d  \eta}\,=\,-\Lb 1 - P\Rb \,\Lb 1 - \bar P \Rb \,\bar P;
\eeq
Interestingly, although it is not obvious from the form of the
 Hamiltonian (see Eq.(\ref{HBB})), the evolution has the same fixed 
points as in two transverse dimensions, $(0,0), \ (1,0), \ (0,1), \ 
(1,1)$.

Since the Hamiltonian is conserved, we have 
\beq\label{con}
\bar P P = {\rm Const} \equiv {\rm \alpha}
\eeq

Using this conservation relation we obtain 
 the following equations of motion:
 \beq \label{H011}
 \frac{ d P}{d  \eta}\,=\,\Lb  P\,-\,\alpha \Rb \,\Lb 1 - P \Rb ; \ \ \ \ \ \ \frac{ d \bar P}{d  \eta}\,=-\,\Lb  \bar P\,-\,\alpha \Rb \,\Lb 1 - \bar P \Rb 
 \eeq
 It is instructive to note that fixed points $(0,0),\ (0,1)$ and $(1,0)$ 
are not present in these equations, which means that  they are not 
 reachable at $\alpha\ne 0$. The point $(1,1)$ is also not reachable by 
 evolution for $\alpha\ne 1$.
 \eq{H011} has  only two interesting fixed points:
 $\Lb 1,\alpha\Rb$ and $\Lb \alpha, 1\Rb$. Since for any physical
 initial condition $P(0)>\alpha$, the asymptotics at
 $\eta\rightarrow\infty$ is always dominated by the fixed
 point $(P=1,\bar P=\alpha)$, while for $\eta\rightarrow 0$ the
 point $(P=\alpha, \bar P=1)$ is approached.

 The general solution to \eq{H01} takes the form:
 
 \beq \label{H03}
 P (\eta)\,=\,\frac{ \alpha +\beta e^{ (1 - \alpha) \eta} }{1 + \beta e^{ (1 - \alpha)  \eta}}; \ \ \ \ \bar P(\eta)=   \frac{ \alpha (1+\beta e^{ (1 - \alpha) \eta}) }{\alpha +  \beta e^{ (1 - \alpha)  \eta}};
 \eeq
 where the parameters $\beta$ and $\alpha$ should be determined from 
the
 boundary conditions:
 \beq \label{H0BC}
 P (\eta= 0)\,=\,p_0;\,\,\,\,\,\,\,\, \bar P (\eta= Y)\,=\,\frac{\alpha}{P (\eta= Y)}\,=\,\bar p_0
 \eeq
 One can see that for $p_0 \,>\,\bar p_0$ and $e^{(1 - \alpha)Y}\,\gg\,1$  \eq{H0BC} leads to
 \beq \label{H031}
 \beta\,=\,\frac{p_0 \,-\,\alpha}{1\,-\,p_0}\,=\,\frac{p_0 \,-\,\bar p_0}{1 - p_0}; \,\,\,\,\,\,\,\,\,\,\alpha\,=\,\bar p_0;
 \eeq 
 For  a symmetric boundary condition $p_0 = \bar p_0$, \eq{H0BC} gives
 $P(0)=\bar P(Y)$, and the solution takes the form

\begin{eqnarray}\label{H032}
P(\eta)&=&\frac{\alpha+\sqrt{\alpha}e^{(1-\alpha)(\eta-Y/2)}}{1+\sqrt{\alpha}e^{(1-\alpha)(\eta-Y/2)}}\\
\bar P(\eta)&=&\frac{\alpha\left(1+\sqrt{\alpha}e^{(1-\alpha)(\eta-Y/2)}\right)}{\alpha+\sqrt{\alpha}e^{(1-\alpha)(\eta-Y/2)}}\nonumber\\
P(\eta)&=&\bar P(Y-\eta)\nonumber
\end{eqnarray}

This solution has a distinct   BFKL limit.  Indeed,
for  $\alpha\ll 1$ and $e^{-Y/2}=a\sqrt{\alpha}$ with
 $1/\alpha\gg a\gg 1$. 

We now have  the BFKL-like contribution
\beq \label{POM}
P(\eta)\,\,\approx \,\,a\alpha e^{\eta}\eeq

Comparing \eq{POM} with \eq{TM1} and \eq{N} one can see that
 the variable $Y$ in all our formulae is actually equal  to $\Delta Y$.
 The exponential ``BFKL-like'' growth continues until the Pomeron
 reaches the value $P(Y)=1$.

The classical solutions determine the scattering amplitude in the
 classical approximation.  The scattering amplitude has a path
 integral representation. The Pomeron Lagrangian that generates
 the equations of motion eq.(\ref{H01}) is
\beq
{\cal L}^{\rm NPM}=\int_0^Yd\eta\left[\frac{1}{\gamma} \ln(1-P)\frac{\partial}{\partial \eta}\ln (1-\bar P) -H\right]= \frac{1}{\gamma} \int_0^Yd\eta\left[ \ln(1-P)\frac{\partial}{\partial \eta}\ln (1-\bar P) 
+\bar PP\right]
\eeq
The scattering amplitude is then given by 
\beq
S^{\rm NPM}_{m\bar n}(Y)=\int dP(\eta)d\bar P(\eta)e^{\frac{1}{\gamma} \int_0^Yd\eta\left[ \ln(1-P)\frac{\partial}{\partial \eta}\ln (1-\bar P) 
+\bar PP\right]}(1-P(Y))^m(1-\bar P(0))^{\bar n}
\eeq
In the classical approximation
\begin{eqnarray}\label{classs}
S^{\rm NPM}_{m\bar n}(Y)&=&e^{\frac{1}{\gamma} \int_0^Yd\eta\left[ \ln(1-p)\frac{\partial}{\partial \eta}\ln (1-\bar p) 
+\bar pp\right]}[1-p(Y)]^m[1-\bar p(0)]^{\bar n}|_{p(0)=1-e^{-\gamma \bar n};\  \bar p(Y)=1-e^{-\gamma m}}\nonumber\\
&=&[1-p(Y)]^m\,e^{\frac{1}{\gamma}\int_0^Yd\eta \left[\ln(1-\bar p)+\bar p\right]p}
\end{eqnarray}
where $p(\eta)$ and $\bar p(\eta)$ denote the solutions of the classical 
equations
 of motion, with the boundary conditions specified in \eq{classs}).

It is interesting to compare the scattering amplitude given by this
 expression, to the one obtained from the BK equation, which in QCD  
describes
  deep inelastic scattering with nuclei. For the latter we have
\beq
S^{\rm BK}_{m\bar n}(Y)=\int dP(\eta)d\bar P(\eta)e^{\frac{1}{\gamma}
 \int_0^Yd\eta\left[ \ln(1-P)\frac{\partial}{\partial \eta}\ln (1-\bar P) -
\ln (1-\bar P)PP\right]}(1-P(Y))^m(1-\bar P(0))^{\bar n}
\eeq
In the classical approximation
\begin{eqnarray}\label{classs4}
S^{\rm BK}_{m\bar n}(Y)&=&e^{\frac{1}{\gamma} \int_0^Yd\eta\left[
 \ln(1-p)\frac{\partial}{\partial \eta}\ln (1-\bar p) 
-\ln(1-\bar p)p\right]}[1-p(Y)]^m[1-\bar p(0)]^{\bar n}|_{p(0)=1-e^{-\gamma \bar n};\  \bar p(Y)=1-e^{-\gamma m}}\nonumber\\
&=&[1-p(Y)]^m
\end{eqnarray}
Note that the solution for $\bar P$ is irrelevant for the BK amplitude,
 which is determined entirely by $P(Y)$. On the other hand, the scattering
 amplitude in NPM  depends on $\bar P$. Nevertheless the two models
 should be  approximately the same in the regime where the BK 
evolution applies. The results
 of the estimates in Ref.\cite{KLL} shows that  in the region  close to 
saturation the differences between BK and NPM are quite significant.
 We will continue the comparison of the two approaches in the 
following sections,
 where we construct a realistic model based on the NPM approach.

 %%%%%%%%%%%%%%%%%%%%%%%%%%%%%%%%%%%%%%%%%%%%%%%%%%%%
 \section{The model}
%%%%%%%%%%%%%%%%%%%%%%%%%%%%%%%%%%%%%%%%%%%%%%%%%%%%%%

To build a model we need to solve two problems:(i) to express
 parameters $\alpha$ and $\beta$ in \eq{H03}  in terms of  $p_0$ and 
$\bar{p}_0$,
 and to take the integral over $\eta$ in \eq{classs}; (ii)  to introduce
 the non-perturbative structure of hadrons.

 %%%%%%%%%%%%%%%%%%%%%%%%%%%%%%%%%%%%%%%%%%%%%%%%%%%%
 \subsection{ Explicit solutions}
%%%%%%%%%%%%%%%%%%%%%%%%%%%%%%%%%%%%%%%%%%%%%%%%%%%
We start with solving the first  problem, which although it is technical, 
   simplifies the fitting procedure. In the general case we need
 to replace \eq{H0BC}  and \eq{H031} by the following expressions:

\bea 
\alpha\Lb p, \bar{p}, z_m\Rb &=&\,\h\Lb p_0 + \bar{p}_0\Rb  \,-\,\frac{1}{2\,z_m}\Lb 1 - D\Rb; ~~
\beta\Lb p, \bar{p}, z_m\Rb \,\,=\, \h \frac{p_0 - \bar{p}_0}{1 - p_0} -\frac{1}{2 z_m (1 - p_0)}\Lb (1- p_0) (1 - \bar{p}_0)  -   D\Rb;\label{ALBE}\\
D &=& \sqrt{4 p_0 (1 - p_0) (1-  \bar{p}_0) z_m -  \Lb (1 - p_0) (1-  \bar{p}_0) - (p_0 -  \bar{p}_0) z_m\Rb^2};\label{D}
\eea

 After doing the integration in \eq{classs}  we obtain
:
\beq \label{INT1}
S\Lb \alpha, \beta,z_m\Rb\,\,=\,\frac{1}{\gamma}\int_0^Yd\eta \left[\ln(1-\bar p(\eta))+\bar p(\eta)\right]\,p(\eta) \,\,=\,\,\tilde{S}\Lb \alpha, \beta,z_m\Rb\,\,-\,\,
\tilde{S}\Lb \alpha, \beta,z_m=1\Rb
\eeq
with
\bea
\tilde{S}\Lb \alpha, \beta,z_m\Rb &=&-(\alpha-1) \text{Li}_2(-\beta z_m)+\alpha
   \text{Li}_2\left(-\frac{\beta
   z_m}{\alpha}\right)+(\alpha-1)
   \text{Li}_2\left(\frac{\alpha+\beta
   z_m}{\alpha-1}\right)+\frac{1}{2} \alpha \ln
   ^2((1-\alpha) \beta z_m)\nn\\
   & &-(\alpha-1) \ln (\beta z_m +1)
   \ln ((1-\alpha) \beta z_m)-\left(\alpha \ln
   (z_m)-(\alpha-1) \ln \left(-\frac{\beta
   z_m+1}{\alpha-1}\right)\right) \ln (\alpha+ \beta
   z_m)\nn\\
    & &+\alpha \ln (z_m) \ln \left(\frac{\beta
   z_m}{\alpha}+1\right)
\eea
where $z_m = e^{\Delta(1 - p_0)Y}$.

 %%%%%%%%%%%%%%%%%%%%%%%%%%%%%%%%%%%%%%%%%%%%%%%%%%%
 \subsection{ Non-perturbative structure of hadrons}
 %%%%%%%%%%%%%%%%%%%%%%%%%%%%%%%%%%%%%%%%%%%%%%%%%%%
In this paper we only  discuss  the non-perturbative 
corrections,
 which are related to the impact parameter dependence of the scattering
 amplitudes. Our assumptions  regarding the hadron structure are 
based on the
 following features of the Pomeron interactions which stem from the
 Colour Glass Condensate (CGC/saturation) effective QCD theory at
 high energies (see Ref.\cite{KOLEB} for a review):
\begin{enumerate}
\item \quad The BFKL Pomeron exchange occurs at fixed impact parameters.
 In other words the Green function of the Pomeron $\propto \,\delta^{(2)}\Lb
 \vec{b}\Rb$.  The triple Pomeron vertex (see \fig{gen})  does not  change 
the impact parameters. We have discussed this property in the introduction.

\item \quad For nucleus-nucleus collisions the main contributions stems from
 the 'net' diagrams of \fig{gen}-a \cite{BRAUN}. In these diagrams the
 dependence on the impact parameters are concentrated in the vertices
 of the Pomeron interaction with the nuclei, this dependence has a
 clear meaning: i.e. the number of nucleons  in a nucleus
 at fixed impact parameter.
\item \quad For  DIS we have the following formula for the total cross section:

\beq\label{FORMULA}
N\Lb Q, Y; b\Rb \,\,=\,\,\int \frac{d^2 r}{4\,\pi} \int^1_0 d z \,
\Psi_{\gamma^*}\Lb Q, r, z\Rb \,N\Lb r, Y; b\Rb\,\Psi^*_V\Lb r,z\Rb
\eeq
where $Y \,=\,\ln\Lb 1/x_{Bj}\Rb$ and $x_{Bj}$ is the Bjorken $x$. $z$
 is the fraction of energy carried by quark.
$Q$ denotes the photon virtuality.
$ N\Lb r, Y; b\Rb $
  is the scattering amplitude for a dipole of size $r$ at 
 impact parameter $b$
which  is the solution to the Balitsky-Kovchegov equation\cite{BK}.
$\Psi_{\gamma^*}\Lb Q, r, z\Rb$ is the wave function of the dipole in 
the  virtual photon.

\eq{FORMULA} splits the calculation of the scattering amplitude into
 two steps: (i) calculation of the wave functions, and (ii) estimates of 
the
 dipole scattering amplitude.

\item \quad The initial condition for the amplitude $N\Lb r, Y; b\Rb$
 can be written as follows \cite{GLA,MUGLA,MV}
\beq \label{GLAMU}
N\Lb r, Y=Y_0 ,b\Rb\,\,=\,\,i \Bigg( 1 \,\,-\,\,\exp\Big( - N^{\rm BA}\Lb r, Q_T=0\Rb \,T_A\Lb b \Rb\Big)\Bigg)
\eeq
where $N^{\rm BA}\Lb r,Q_T\Rb$ is the  amplitude for  dipole scattering
 in the Borm approximation . $T\Lb b \Rb$ denotes the number of the 
nucleons
 at fixed impact parameter $b$.

\end{enumerate}

In \eq{GLAMU},  $ N^{\rm BA}\Lb r, Q_T=0\Rb \,=\,\frac{\as \pi }{2 N_c}
 \pi r^2 \ln\Lb \frac{r^2}{R_N^2}\Rb$ while
 $T\Lb b\Rb \propto 1/(\pi R^2_A)$ ($R_N$ and $R_A$ are the  radii of
  nucleon and nucleus, respectively). Therefore, we can re-write 
 $N^{\rm BA}\Lb r, Q_T=0\Rb \,T_A\Lb b \Rb$ in \eq{GLAMU} as
 $\gamma\,S\Lb b \Rb$ where $\gamma$ is the dipole scattering
 amplitude which is equal to
 $\frac{\as \pi }{2 N_c} \ln\Lb \frac{r^2}{R_N^2}\Rb$, and $S_A\Lb b \Rb$ 
denotes
 the number of the dipoles with size $r$ that 
we have in the nucleus at    impact parameter $b$. \eq{GLAMU} takes
 the form
\beq \label{GLAMU1}
N\Lb r, Y=Y_0 ,b\Rb\,\,=\,\,i \Bigg( 1 \,\,-\,\,\exp\Big( - \gamma^{\rm BA}\Lb r\Rb \,S_A\Lb b \Rb\Big)\Bigg)
\eeq
Comparing \eq{GLAMU1} with \eq{classs} we see that we can interpret the
 initial condition $p(0)=1 - \exp\Lb - \gamma n\Rb$  as $\gamma = 
 \gamma^{\rm BA}\Lb r\Rb$ and $n = S_A\Lb b \Rb$.

Bearing this in mind, we introduce the following initial conditions
 for $p(y)$ and $\bar{p}(y)$, considering $\gamma S_A(b) \,\ll\,1$:

\beq \label{IC}
P\Lb y =y_0, b'\Rb\,\,=\,\,p_0 \,S\Lb b',m\Rb; ~~~~~~\bar{P}\Lb y = Y,\vec{b} - \vec{b}' \Rb\,\,=\,\,p_0 \,S\Lb \vec{b} - \vec{b}',m\Rb;
\eeq
choosing  
\beq \label{S}
S\Lb b, m\Rb\,\,=\,\, m\, b\, K_1(m\, b)
\eeq
where $K_1$ is the McDonald function (see Ref.\cite{RY} formula
 {\bf 8.43}). Comparing \eq{IC} with \eq{GLAMU1} we see that $\gamma = p_0$.

  %%%%%%%%%%%%%%%%%%%%%%%%%%%%%%%%%%%%%%%%%%%%%%%%%%%%%%%%%%%
       \begin{figure}[ht]
    \centering
  \leavevmode
      \includegraphics[width=10cm]{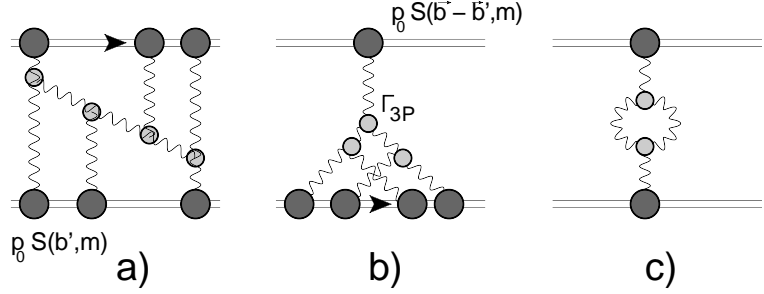}  
      \caption{ Examples of the Pomeron diagrams: 'net' diagrams,
 which give the main contribution to nucleus-nucleus collisions
 (\fig{gen}-a); the diagrams of the Balitsky-Kovchegov equation
 (\fig{gen}-b); and the contribution to the Green function of the
 Pomeron(\fig{gen}-c). The wavy lines describe the Pomeron Green function.}
\label{gen} 
   \end{figure}

 %%%%%%%%%%%%%%%%%%%%%%%%%%%%%%%%%%%%%%%%%%%%%%%%%%%%%%%%%%%%%%%%
The natural generalization of \eq{FORMULA} for diagram of \fig{gen}-b is 
\beq \label{FORMULA1}
N_{\fig{gen}-b}\Lb Y, b\Rb\,\,=\,\,\frac{m^2}{4 \pi}\int d^2 b'\bar{P}\Lb y=Y,\vec{b} - \vec{b}'\Rb\Bigg( 1 - ( 1 - P\Lb Y,b'\Rb)^n\Bigg)
\eeq
which at $Y \to 0$ has the form
\beq \label{FORMULA2}
N_{\fig{gen}-b}\Lb Y=Y_0, b\Rb\,\,=\,\,\frac{m^2}{4 \pi}\int d^2 b' \,\,p^2_0\,\, S\Lb \vec{b} - \vec{b}', m\Rb\,S\Lb  \vec{b}', m\Rb\,\,=\,\,p^2_0\, \,n
\eeq
where $n =\,\frac{m^2}{4 \pi}\int d^2 b\, S\Lb \vec{b} - \vec{b}', m\Rb\,S\Lb
  \vec{b}', m\Rb$ is the number of  dipoles of  size $1/m$ 
 in the two interacting nucleons. It should be stressed that in 
\eq{FORMULA2} we
 assumed that the typical size of the dipole that is described by the new
 parton model is $2/m$. In principle, we can introduce this size as a new
 parameter whose value we will need to determine from comparison with the
 experimental data.

The expression for $z_m = e^{\Delta(1 - p_0)Y}$  sums  diagrams of the 
type
 shown in  \fig{gen}-c . Since the triple Pomeron interaction does not
 induce any impact parameter dependence, we consider
 $p_0 = \gamma = {\rm Const}$ as a function of $b$.

Combining  all the above, we obtain the following equation for the
 scattering amplitude:

\beq
A_{el}(Y, b)\,=\, 1  - \exp \Bigg\{\int \frac{m^2 d^2 b'}{ 4 \pi} \Bigg[\frac{1}{p_0} \Big( S(\alpha,\beta,z_m) \,{\mathbf +}\, \alpha(p, \bar{p},z_m) \Delta (1 - p_0) Y\Big)  \,-\,\bar{P}\Lb y = Y, \vec{b} - \vec{b}'\Rb\, P\Lb Y, b'\Rb\Bigg]\Bigg\}
\eeq

 %%%%%%%%%%%%%%%%%%%%%%%%%%%%%%%%%%%%%%%%%%%%%%%%%%%
 \section{Comparison with   experimental data}
%%%%%%%%%%%%%%%%%%%%%%%%%%%%%%%%%%%%%%%%%%%%%%%%%%%

 %%%%%%%%%%%%%%%%%%%%%%%%%%%%%%%%%%%%%%%%%%%%%%%%%%%
 \begin{boldmath}
 \subsection{Fitting $\sigma_{\rm tot}, \sigma_{\rm el}$ and $B_{\rm el}$}
 \end{boldmath}
%%%%%%%%%%%%%%%%%%%%%%%%%%%%%%%%%%%%%%%%%%%%%%%%%%%

As we have seen in the previous section, we introduce two dimensionless
 parameters:    $\Delta$ - the intercept of the BFKL Pomeron, and $\gamma$ -
 the amplitude  for dipole-dipole scattering at low energies. For
 $b$-dependence we suggested a specific form for $b$-dependence (see \eq{IC})
 which is characterized by the dimensional factor $m$. All  three
 parameters  were determined by  fitting to  the experimental data. 
We
 choose to describe three observables: total  and elastic cross section
 and the elastic slope. They have the following expressions through the
 partial amplitudes:
\bea \label{OBS}
 \sigma_{tot}\,&=&\,\,2\,\int d^2 b \,\mbox{Im} A_{el}(Y, b); \nn\\
  \sigma_{el}\,&=&\,\,\,\int d^2 b \,| A_{el}(Y, b)|^2;\nn\\
 B_{el}\,&=&\, \frac{1}{2} \frac{\int b^2\, d^2 b \,\mbox{Im} A_{el}(Y, b)}{\int d^2 b \,\mbox{Im} A_{el}(Y,  b)};\eea
 
 From \fig{fit} one can see that we can describe the data for $W \geq 1\,TeV$.
 The values of parameters are shown in Table 1. Comparing these parameters with
 the resulting curves in \fig{fit} we  note that the shadowing 
corrections play an
  essential role. First, the corrections to the Green function of the 
Pomeron reduce the Pomeron intercept from $\Delta = 0.644$ to $\Delta_{\rm
 dresssed}\,\,=\,\,0.35$. The other shadowing corrections lead to the
 effective intercept $\Delta_{\rm eff} \approx 0.07$.

%%%%%%%%%%%%%%%%%%%%%%%%%%%%%%%%%%%%%%%%%%%%%%%%%%%%%%%%%%%
\begin{table}[h]
\begin{minipage}{10cm}{
\begin{tabular}{|l|l|l|l|}
\hline
$\Delta$ & $p_0$  &m (GeV) &$\chi^2$/d.o.f.\\
\hline
 0.6488 $\pm$ 0.030 &0.489 $\pm$ 0.030 &0.867  $\pm$ 0.005& 1.3  \\\hline
\end{tabular}
}
\end{minipage}
\begin{minipage}{5cm}
{\caption{Fitted parameters.}}
\end{minipage}
\label{t2}
\end{table}
%%%%%%%%%%%%%%%%%%%%%%%%%%%%%%%%%%%%%%%%%%%%%%%%%%  
From \fig{fit} we see that we  fail to describe  the 
experimental data for $ W  < 1 \, TeV $.
  However, we would like to stress that we used a very naive model for the
 hadron structure. Our previous experience\cite{GLMNI,GLM2CH} shows that we
 need to take into account the processes of diffraction production, 
which
 have been neglected in this model. Our main goal 
is to demonstrate that the suggested model is able to describe  the 
experimental data at  high energies. 
 
 %%%%%%%%%%%%%%%%%%%%%%%%%%%%%%%%%%%%%%%%%%%%%%%%%%%%%%%%%%%%%%%%
       \begin{figure}[ht]
    \centering
  \leavevmode
  \begin{tabular}{c c c}
      \includegraphics[width=6cm]{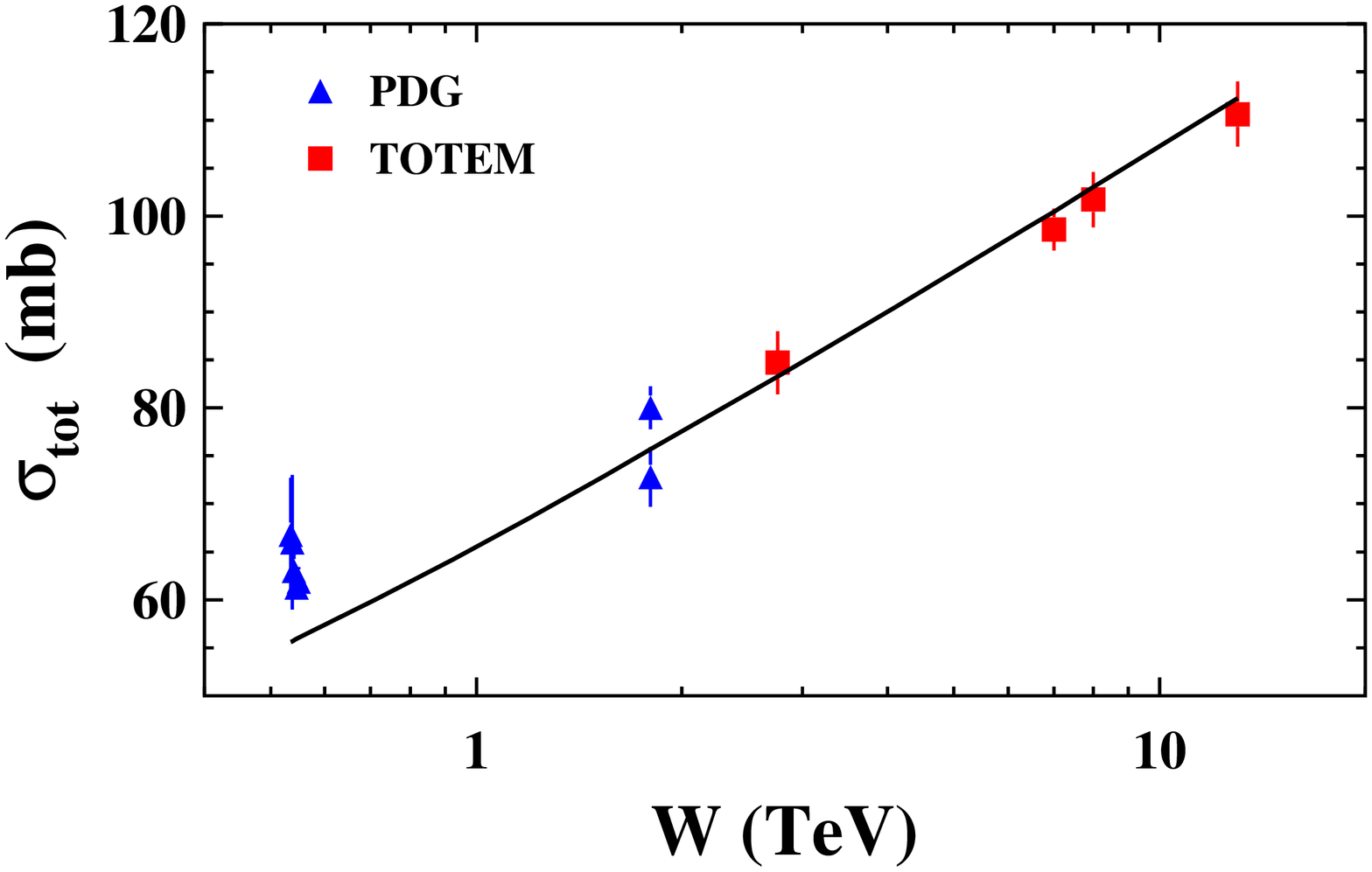}  &\includegraphics[width=6cm]{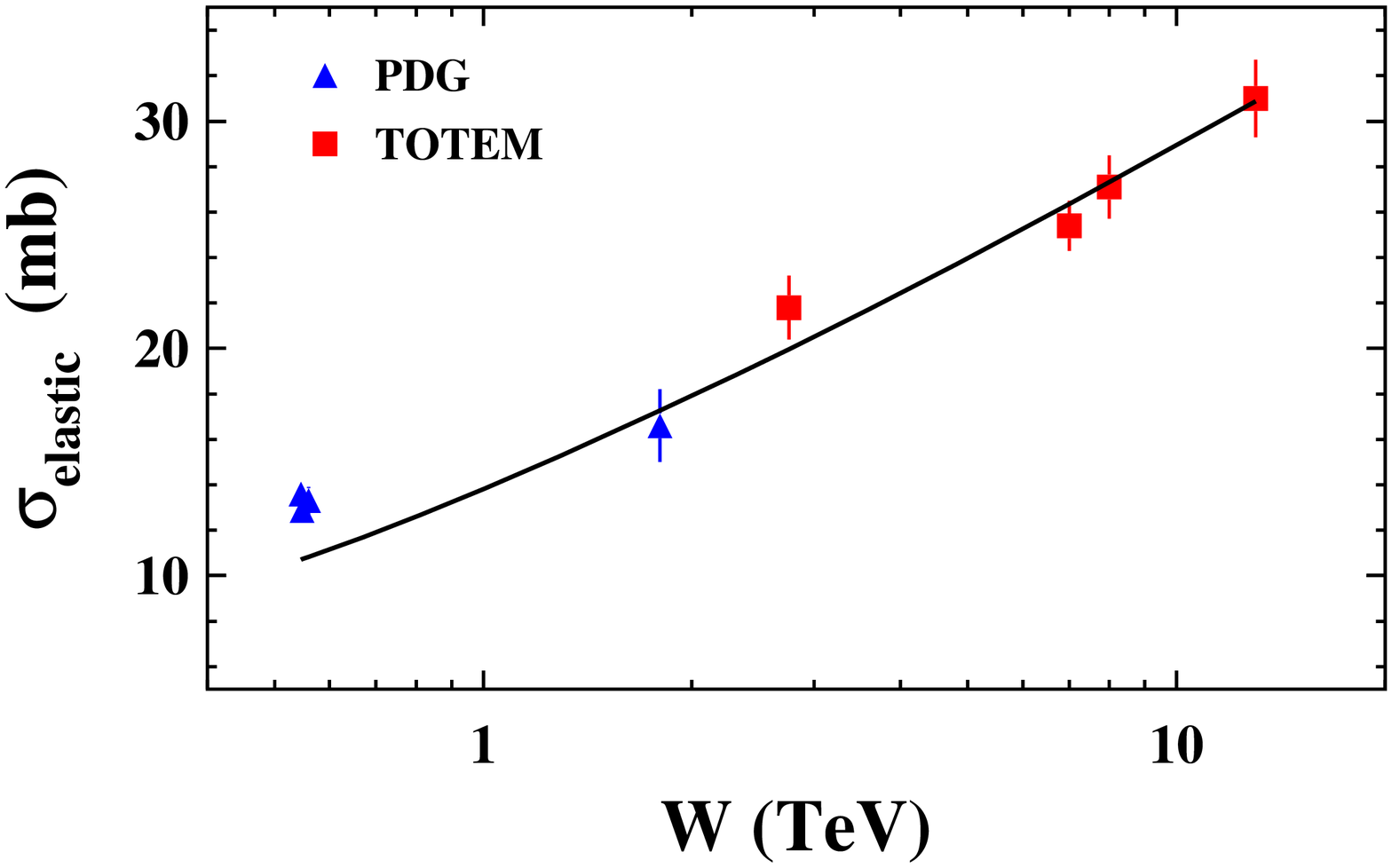}&\includegraphics[width=6cm]{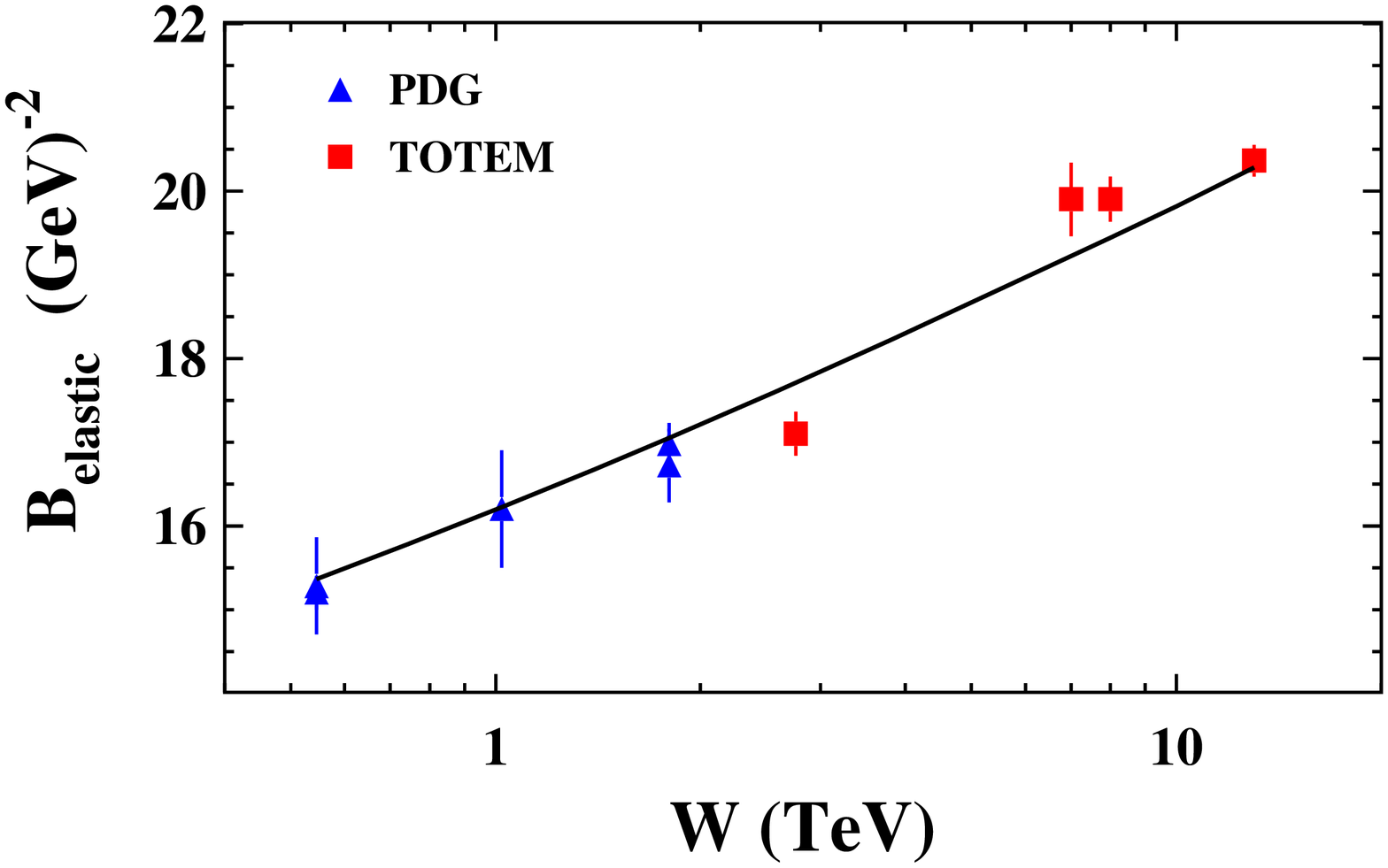} \\
      \fig{fit}-a&\fig{fit}-b& \fig{fit}-c\\
      \end{tabular}
      
 \caption{ The energy behaviour of $\sigma_{tot},\sigma_{el}$ and the
 slope $B_{el}$ for proton-proton scattering  as predicted in our 
model. 
 Data are taken from Refs.\cite{PDG,TOTEMRHO}.  The data taken for the
 fit were for $W \,\geq\,$ 1 TeV.  }
 \label{fit}
   \end{figure}

 %%%%%%%%%%%%%%%%%%%%%%%%%%%%%%%%%%%%%%%%%%%%%%%%%%%%%%%%%%%%%%%   

 %%%%%%%%%%%%%%%%%%%%%%%%%%%%%%%%%%%%%%%%%%%%%%%%%%%
\begin{boldmath}
 \subsection{$A_{el}(Y,  b)$}
\end{boldmath}
%%%%%%%%%%%%%%%%%%%%%%%%%%%%%%%%%%%%%%%%%%%%%%%%%%%

In \fig{ael} we plot the elastic scattering amplitude 
$A_{\rm el}\Lb Y, b\Rb$, as a function of the 
 impact parameter.  Note  that this amplitude
 has reached the unitary limit 1 at $W=13 \,TeV$ and shows the increasing with
  energy  radius, of the interaction. For a comparison,  we include in 
this
 picture the scattering amplitude, calculated in Balitsky-Kovchegov
 non-linear equation (see $A_{\rm el}(X)$). This amplitude is far 
 from the unitarity limit and   shows only  a small  increment with 
  increasing energy.

  %%%%%%%%%%%%%%%%%%%%%%%%%%%%%%%%%%%%%%%%%%%%%%%%%%%%%%%%%%%
       \begin{figure}[ht]
    \centering
  \leavevmode
      \includegraphics[width=8cm]{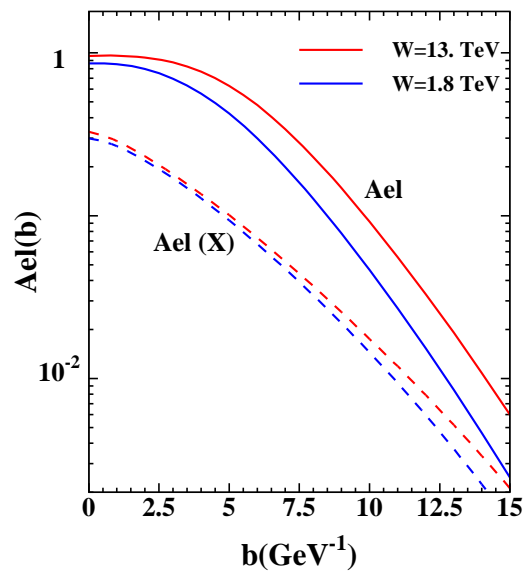}  
      \caption{ The elastic scattering amplitude versus $b$.
 $A_{\rm el}(X)$ is the amplitude from the Balitsky-Kovchegov
 equation( see \eq{classs4}).}
      
\label{ael} 
   \end{figure}

 %%%%%%%%%%%%%%%%%%%%%%%%%%%%%%%%%%%%%%%%%%%%%%%%%%%%%%%%%%%%%%%%
Such behaviour of the scattering amplitude  $A_{\rm el}(X)$ reflects
 the fact that in our approach  there is no fixed point (1,0) (or/ and
 (0,1)), as  occurs  in the Braun Hamiltonian, and the scattering of 
hadrons
 or nucleus cannot be reduced to BK evolution at high energies.

 ~

%%%%%%%%%%%%%%%%%%%%%%%%%%%%%%%%%%%%%%%%%%%%%%%%%%%
 \section{Conclusions}
%%%%%%%%%%%%%%%%%%%%%%%%%%%%%%%%%%%%%%%%%%%%%%%%%%%

In this paper we showed that the experimental data at high energies, can 
be
 described in the framework of the new parton model. The model is based on
 the Pomeron calculus in 1+1 space-time, suggested in Ref. \cite{AKLL}, and
 on  simple assumptions on the hadron structure, related to the impact
 parameter dependence of the scattering amplitude. This parton model stems 
 from QCD, assuming that the unknown non-perturbative corrections lead to 
fixing  the size of the interacting dipoles. The advantage of this approach
 is that it satisfies both the t-channel and s-channel unitarity, and can 
be 
used for summing all diagrams of the Pomeron interaction, including 
 Pomeron loops. In other words, we can use this approach for all possible
 reactions: dilute-dilute (hadron-hadron), dilute-dense (hadron - nucleus)
 and dense-dense (nucleus-nucleus) parton systems scattering.  Unfortunately,
 we are still far  from   tackling this problem in the framework 
of QCD effective theory at
 high energy (CGC /saturation approach).

We achieved quite  good descriptions of the three experimental 
observables:
 $\sigma_{\rm tot}$,$\sigma_{\rm el}$
and $B_{\rm el}$, especially  regarding the energy dependence
 of these observables.  We consider this paper as the first attempt
 to show that the new parton model  can be relevant to the discussion of the
 experimental data. In spite of the embryonic state of the theory of the
 quark-gluon confinement, we  hope that our model can be viewed as the
 first step in the right direction regarding the theoretical 
description of the
 dilute-dilute parton system scattering at high energy at which, we
 believe,
 that such systems become  dense.

We are  aware that our model is very naive in the description of the
 hadron structure. We are planning to include  diffraction production
 in our formalism, and to develop a theoretical approach in the 
framework
 of the new parton model, so as to be able  to treat  processes of  
multiparticle  generation. 

~

 {\it Acknowledgements.} \\
   We thank our colleagues at Tel Aviv University and UTFSM for
 encouraging discussions.
 This research was supported by the BSF grant   2012124, by 
   Proyecto Basal FB 0821(Chile) ,  Fondecyt (Chile) grants  
 1140842 and 1180118 and by   CONICYT grant PIA ACT1406.

\end{document}